\documentclass[onecolumn]{article} % Сразу задаем один столбец

\usepackage{microtype}
\usepackage{graphicx}
\usepackage{subfigure}
\usepackage{booktabs} % for professional tables
\usepackage{dsfont}
\usepackage{soul, xcolor}
\usepackage{float}
\usepackage{tabularx}
\usepackage{caption}
\usepackage[utf8]{inputenc}
\usepackage[T1]{fontenc}

\usepackage{hyperref}

\usepackage[accepted]{icml2025}

\usepackage{amsmath}
\usepackage{amssymb}
\usepackage{mathtools}
\usepackage{amsthm}

\usepackage[capitalize,noabbrev]{cleveref}

\theoremstyle{plain}

\theoremstyle{definition}

\theoremstyle{remark}

\usepackage[textsize=tiny]{todonotes}

\begin{document}

\onecolumn 

\begin{center}
    \vskip 0.25in
    \begin{center}
        {\huge \bf DCD: Domain-Oriented Design for Controlled \\ Retrieval-Augmented Generation \par}
    \end{center}
    \vskip 0.1in
    \hrule height 1pt 
    \vskip 0.3in
    
    {\large Valerii Kovalskii, Nikita Belov, Nikita Miteyko, Igor Reshetnikov, Maksim Maksimov} \\
    
\end{center}

\begin{center} \textbf{red\_mad\_robot} \end{center}


\vskip 0.3in



\begin{abstract}
Retrieval-Augmented Generation (RAG) is widely used to ground large language models in external knowledge sources. However, when applied to heterogeneous corpora and multi-step queries, Naive RAG pipelines often degrade in quality due to flat knowledge representations and the absence of explicit workflows.
In this work, we introduce DCD (Domain–Collection–Document), a domain-oriented design to structure knowledge and control query processing in RAG systems without modifying the underlying language model. The proposed approach relies on a hierarchical decomposition of the information space and multi-stage routing based on structured model outputs, enabling progressive restriction of both retrieval and generation scopes.
The architecture is complemented by smart chunking, hybrid retrieval, and integrated validation and generation guardrail mechanisms. We describe the DCD architecture and workflow and discuss evaluation results on synthetic evaluation dataset, highlighting their impact on robustness, factual accuracy, and answer relevance in applied RAG scenarios.

\end{abstract}

\section{Introduction}
\label{submission}
RAG has gained widespread adoption as a practical approach for integrating language models with external knowledge sources [\hyperlink{Lewis}{Lewis et al., 2020}]. Even basic Naive RAG implementations can effectively address a broad range of applied tasks, from customer support query handling to enterprise document analysis [\hyperlink{Izacard}{Izacard et al., 2021}]. 

However, as data volumes grow and user queries become more complex, the limitations of standard RAG pipelines become increasingly apparent. In particular, the linear architecture of Naive RAG exhibits degraded performance on multi-step queries that require sequential interpretation, retrieval across multiple knowledge slices, and result aggregation [\hyperlink{Wei}{Wei et al., 2023}]. Another critical issue is the sensitivity of RAG systems to document segmentation parameters. These factors limit the applicability of Naive RAG in scenarios demanding high accuracy and predictable behavior.

Existing approaches often improve RAG systems through additional reasoning loops or agent orchestration layers. In contrast, this work follows an architectural rather than agentic direction. We propose DCD (Domain–Collection–Document), a domain-oriented RAG design that improves retrieval robustness through an explicit hierarchical organization of knowledge and a controlled query-processing workflow. Instead of relying on additional reasoning mechanisms, DCD constrains retrieval through hierarchical decomposition of the knowledge base. The proposed approach improves answer quality for multi-step queries while preserving reproducibility, controllability, and operational safety without modifying the underlying language models.

\section{Preliminaries}
\subsection{Retrieval-Augmented Generation}
We focus on improving the accuracy and robustness of RAG systems in scenarios involving multi-step queries and heterogeneous knowledge corpora. Specifically, we consider architectural approaches that enable:
\begin{itemize}
\item restricting the retrieval space to relevant subsets of knowledge,
\item explicit control over query processing workflows,
\item reproducibility and quality assurance of generated outputs.
\end{itemize}

Throughout this work, Naive RAG refers to a linear pipeline without explicit knowledge structuring, query decomposition, or multi-stage generation control. We deliberately focus not on modifying the language model itself, but on organizing knowledge and workflows around it.

In many practical RAG implementations, retrieval is performed once at the beginning of the pipeline and remains fixed throughout the generation process. This design assumes that a single retrieval step can provide all information required to answer the query. However, when the knowledge corpus contains documents from multiple semantic domains, similarity-based retrieval may return fragments that are locally relevant but belong to different knowledge contexts. As a result, the retrieved context may contain partially compatible evidence, forcing the language model to reconcile inconsistent information during generation [\hyperlink{Li}{Li et al., 2025}].

We conceptualize a RAG system as a composition of the following stages:
\begin{itemize}
\item user query interpretation,
\item selection of a relevant knowledge subset,
\item retrieval and aggregation of contextual information,
\item answer generation by a language model,
\item validation and quality control of the output [\hyperlink{Guu}{Guu et al., 2020}].

\end{itemize}

In standard implementations, these stages are often implicit or collapsed into a single step. In contrast, our approach treats each stage as an explicit workflow component, enabling fine-grained control and additional validation.

\subsection{Chunking}
The quality of a RAG system is highly dependent on document segmentation strategies. Common chunking approaches include paragraph-based segmentation, fixed-length chunks, and sliding-window techniques [\hyperlink{Gao}{Gao et al., 2021}].
In our implementation, we apply a sliding-window segmentation strategy with overlap ranging from 5–10\% of the chunk size. A 5\% overlap is typically sufficient to preserve contextual continuity between adjacent chunks. For documents containing semantically similar or repetitive sections, the overlap is increased to 10\% to better capture contextual dependencies.

\subsection{Guardrails}
Classical RAG implementations typically rely on built-in safety mechanisms of language models and lack explicit architectural procedures for validating generated outputs [\hyperlink{Ji}{Ji et al., 2023}]. In particular, they provide no systematic means to verify alignment between generated answers and retrieved context or to refuse generation outside the system’s functional scope [\hyperlink{Lin} {Lin et al., 2022}].
Our system incorporates a guardrails mechanism combined with a hallucination prevention module. This mechanism consists of two interrelated components.
The first component analyzes user queries for prohibited topics using a composite dictionary that includes both a base set of stop-words and custom elements defined by the system owner.
The second component analyzes generated outputs, ensuring the absence of prohibited content and detecting hallucinations. This is achieved by comparing the original query, the retrieved context, and the generated response, enabling early detection of deviations during generation.

\section{Method}
\subsection{Key Assumption}
The central methodological assumption of this work is that answer quality significantly improves when retrieval and generation are constrained to semantically homogeneous knowledge regions — a subset of the corpus whose documents share a common topical scope, terminology, and expected user intent, while remaining clearly distinguishable from documents belonging to other regions [\hyperlink{Yao}{Yao et al., 2023}]. Within such a region, documents address closely related subject areas and therefore compete primarily with semantically relevant sources during retrieval.

This constraint allows the retrieval process to operate within a bounded semantic space, reducing interference from unrelated documents and improving the alignment between retrieved evidence and the informational structure of the query. Rather than operating over a monolithic document store, we assume a prior decomposition of the corpus into independent subspaces, within which documents compete only with semantically similar sources [\hyperlink{Khot}{Khot et al., 2023}].

This issue is closely related to the organization of the knowledge space itself. When retrieval operates over a heterogeneous corpus without explicit semantic boundaries, documents from different knowledge regions compete within the same search space. In such settings, embedding similarity alone may not reflect the structural dependencies between pieces of information required to answer the query. As a result, retrieved evidence may be individually relevant but collectively insufficient for coherent reasoning [\hyperlink{Li}{Li et al., 2025}].

Evidence retrieval studies for long documents demonstrate that coarse-to-fine search strategies can improve the identification of relevant evidence compared to flat fragment-level retrieval. In such approaches, higher-level structural segments are first identified and only then used to guide fine-grained passage selection. By narrowing the search space to semantically coherent regions of a document, retrieval systems preserve global contextual relationships and improve the reliability of downstream question answering [\hyperlink {Nair} {Nair et al., 2023}].

\begin{figure}
    \centering
    \includegraphics[width=1\linewidth]{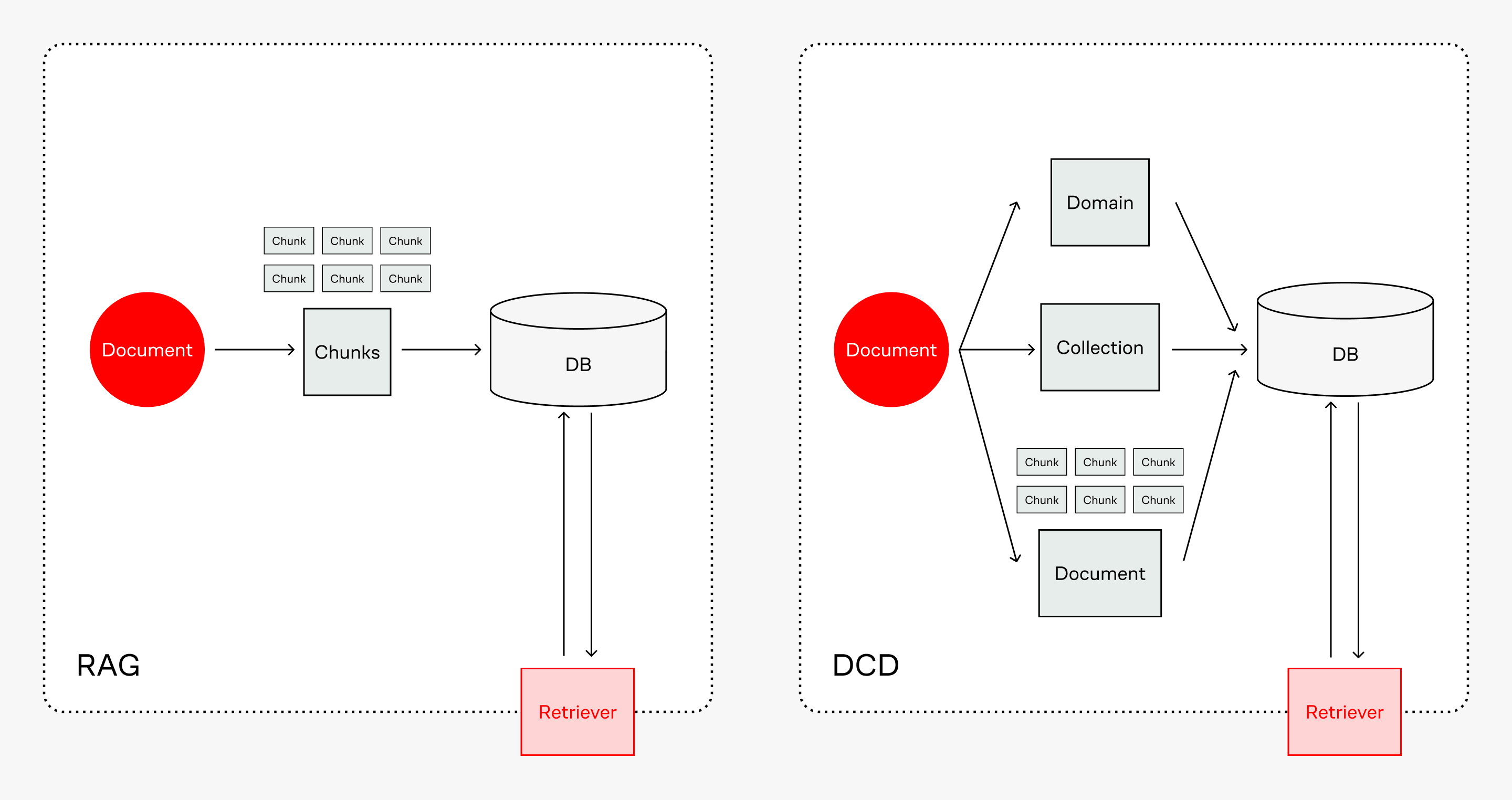}
    \caption{The difference between the RAG and DCD approaches}
    \label{fig:placeholder1}
\end{figure}

The proposed DCD architecture addresses this limitation through explicit decomposition of the knowledge space prior to retrieval. By constraining retrieval to semantically homogeneous regions, the system reduces cross-domain interference and improves alignment between retrieved context and the reasoning structure of the query.

\subsection{Experimental Validation of the Assumption}

To validate this assumption, we conducted experiments on a synthetic dataset constructed to approximate real-world RAG scenarios. The corpora consist of heterogeneous documents with varying structures and semantic granularity — conditions under which pipelines most frequently degrade due to flat knowledge organization and unstable retrieval.

All datasets were anonymized for confidentiality: client identifiers and sensitive metadata were removed. The paper reports procedures and observed effects without disclosing document contents.
In addition to quality metrics, we recorded operational characteristics of the pipeline, including total evaluation time, mean end-to-end response time, and time to first token, allowing analysis of the trade-off between accuracy and latency under increased workflow complexity.

\section{DCD: Domain–Collection–Document}
The DCD (Domain–Collection–Document) Design is an approach to organizing knowledge in RAG systems through explicit hierarchical segmentation of the information space. Its primary goal is to minimize overlap between knowledge areas and prevent irrelevant context from being passed to the language model [\hyperlink{Makin}{Makin, 2024}].

DCD structures data across three levels:
\begin{itemize}
\item Domain — a high-level subject area,
\item Collection — a thematically homogeneous subset within a domain,
\item Document — an atomic knowledge unit with metadata.
\end{itemize}

\subsection{Hierarchical Levels}
A Domain defines top-level search boundaries and is constructed to minimize semantic overlap with other domains.
A Collection represents a narrower thematic slice within a domain and further restricts the retrieval space. Examples include legal documents, reference materials, or user FAQs within a single domain.
A Document is the basic unit used in the RAG process. Documents are enriched with metadata and may be further segmented into chunks for context retrieval.
Information retrieval follows a top-down principle: from domain to collection to document.

Navigation through the hierarchy is performed by the DCD Router, which sequentially selects the most relevant domain, collection, and document for a given query. At each level, routing is formulated as a selection task over a limited candidate set [\hyperlink{Shazeer}{Shazeer et al., 2017}].
Router decisions are produced using structured outputs from the language model, ensuring transparency, reproducibility, and intermediate result caching. To improve robustness, fallback elements (main domain, main collection, main document) are defined at each level.

\begin{figure*}
    \centering
    \includegraphics[width=1\textwidth]{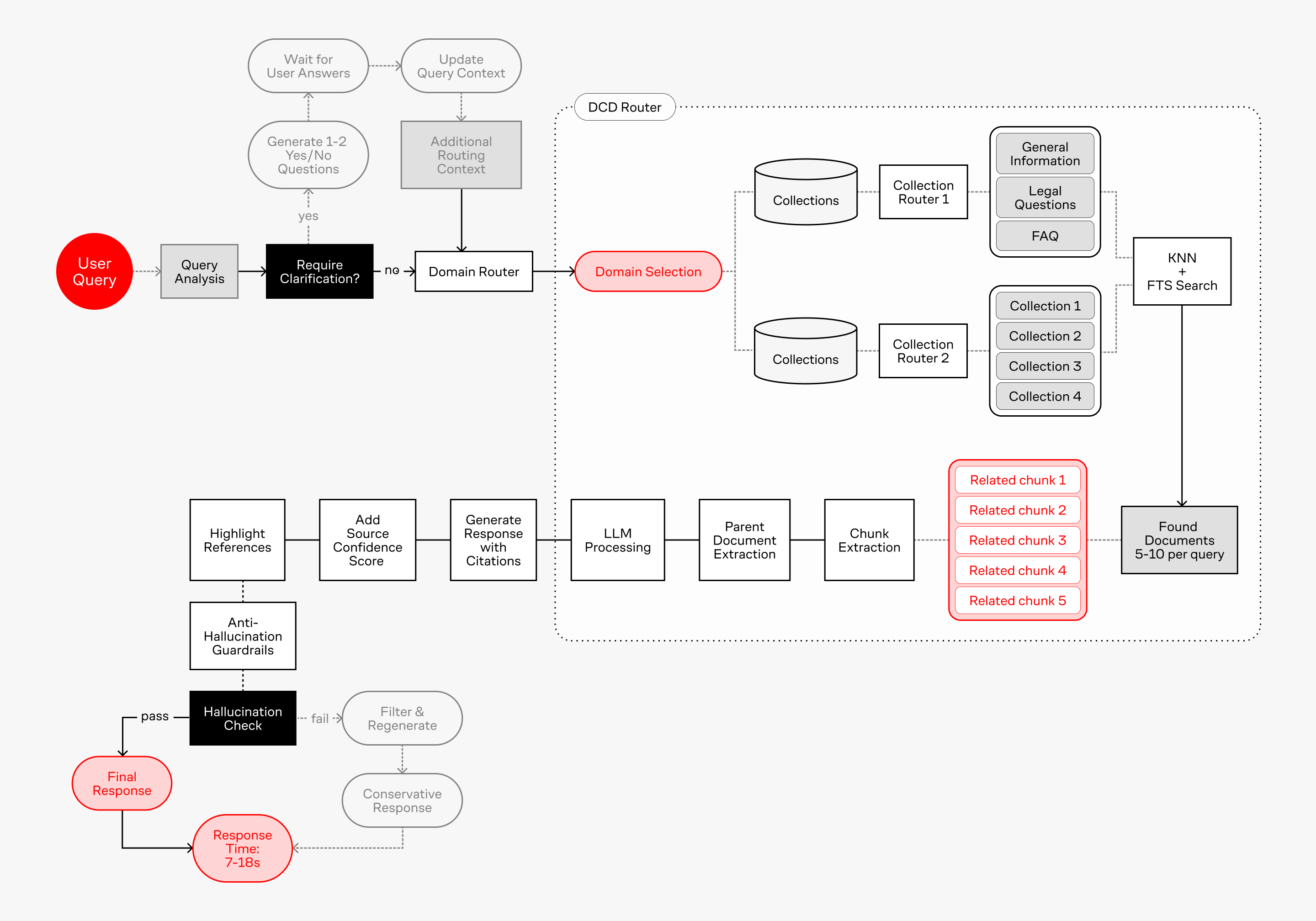}
    \caption {Overview of the DCD retrieval and response generation workflow}
    \label{fig:fullwidth}
\end{figure*}

\subsection{DCD Router}
The functional core of domain segmentation is the DCD Router, a specialized module responsible for identifying the target knowledge segment for retrieval. Like most system components, it relies on structured language model outputs (JSON), enabling high-precision processing and transparent decision-making [\hyperlink{OpenAI}{OpenAI, 2024}].

Initial configuration involves defining domain and collection structures and assigning knowledge assets accordingly. Documents retain their original titles, while domains and collections receive manually defined identifiers. Upon receiving a user query, the system submits it — along with the registry of available domains — to the language model to identify the most relevant domain. This procedure is repeated hierarchically for collections and documents. If the language model produces incorrect outputs, fallback elements at each level ensure system robustness and guarantee a response even after multiple failed routing attempts.

Higher hierarchy levels are described with richer metadata to reduce routing errors, since mistakes at upper levels may propagate downstream and restrict the search space to an incorrect semantic region of the knowledge base. Providing clearer semantic descriptions for domains and collections improves the reliability of the initial routing decisions and helps maintain stable navigation across the hierarchy.

In the present study, routing accuracy at individual hierarchy levels was not evaluated as a separate classification metric. Instead, the routing mechanism is considered as part of the overall retrieval pipeline, and its effectiveness is reflected in downstream system behaviour, including answer relevance, factual accuracy, and retrieval coverage.

\subsection{Integration with the RAG Pipeline}
After a document is selected, the system retrieves relevant fragments using semantic and full-text search, followed by deduplication and reranking. The resulting context is passed to the language model for answer generation.

By constraining the retrieval space upfront, DCD reduces the likelihood of irrelevant context inclusion and improves generation robustness on multi-step queries. While Naive RAG employs a linear processing pipeline — chunking, embedding, retrieval, and generation — DCD preserves a similar user interface but introduces a fundamentally different internal organization based on multi-level processing.

\subsection{Smart Chunking}
Smart chunking is an integral component of the DCD design. Although DCD can operate with classical chunking methods, the selected strategy provided the most stable behaviour within the proposed architecture, as it better preserves the structural properties of the underlying knowledge base.

In this approach, documents are segmented using a sliding window mechanism with partial overlap between neighbouring fragments. This segmentation strategy preserves local semantic continuity and reduces the likelihood that relevant contextual information will be lost at chunk boundaries. As a result, logically connected statements that span multiple sentences or paragraphs remain accessible during retrieval, which is particularly important for queries requiring compositional reasoning.

\begin{figure}[H]
    \centering
    \includegraphics[width=1\linewidth]{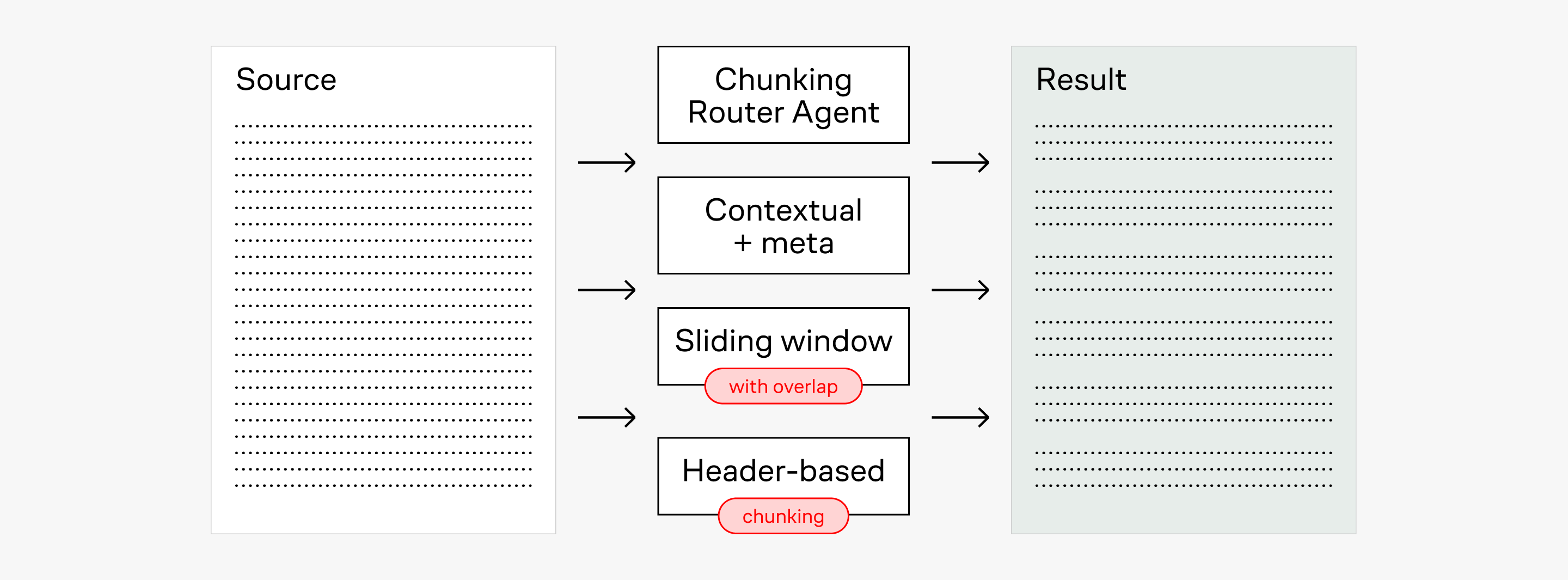}
    \caption{Smart chunking pipeline}
    \label{fig:placeholder2}
\end{figure}

Each fragment is additionally enriched with chunk-level metadata derived from the document structure. These metadata describe the fragment’s position within the document hierarchy and maintain references to the corresponding document-level entity. Such structural annotations allow the retrieval stage to operate not only on semantic similarity but also on contextual relationships between fragments originating from the same document or collection.

The combination of overlapping segmentation and structured metadata improves the consistency of fragment retrieval and reduces the probability of assembling answers from semantically unrelated contexts. This design enables the retrieval pipeline to better align selected evidence with the conceptual structure of the query and contributes to more stable answer generation in retrieval-augmented workflows, as reflected in the experimental evaluation presented later in the paper.

\subsection{Fast Guardrails}
In addition to the main retrieval and generation pipeline, the system introduces a fast guardrails mechanism for early response validation. The system includes a guardrails mechanism combined with hallucination prevention and a fast pre-check function that evaluates responses before full generation completes [\hyperlink{Manakul}{Manakul et al., 2023}].

The guardrails workflow is as follows:

\begin{itemize}
\item the user query undergoes stop-word screening,
\item compliant queries are passed to the DCD RAG pipeline,
\item during response streaming, the first 150 tokens are analyzed,
\item based on query, context, and partial output, the system decides whether to display or block the response.
\end{itemize}

This mechanism filters unsupported queries and validates generated content in parallel with generation, avoiding additional latency.

\subsection{System Completeness Summary}

The DCD architecture demonstrates significant advantages in adaptability when processing heterogeneous data and structurally complex queries. Its flexibility yields substantial accuracy improvements in multi-level knowledge scenarios while integrating critical validation and control stages absent in Naive RAG systems.
While initial configuration requires additional setup effort, these costs are offset by superior performance and rapid amortization through improved operational efficiency.

\section{Metrics}

To comprehensively evaluate the proposed DCD approach, we employ a metric suite extending beyond standard evaluations. The assessment relies on structured generation quality evaluation using an LLM as an assessor [\hyperlink{Liu}{Liu et al., 2023}].

\subsection{Strict Binary Answer Relevance \& Completeness}

$SB ARC$ is a strict binary metric assessing whether an answer is both relevant and complete with respect to a question.
% Requires: \usepackage{amsmath}

\begin{equation}
    S(Q, A) = D(Q, A) \land P(Q, A) \land Sp(Q, A) \land \lnot V(A)
    \label{eq:placeholder_label1}
\end{equation}

$Q, A$ — the question and the answer, respectively,

$S(Q,A)$ — the binary output verdict,

$D(Q,A)$ — the direct-answer criterion,

$P(Q,A)$ — the completeness criterion,

$Sp(Q,A)$ — the specificity criterion,

$V(A)$ — the vagueness criterion,

% Requires: \usepackage{amsmath}
\begin{equation}
    \label{eq:placeholder_label2}
    SB\ ARC = \frac{\sum_{i=1}^{N} S(Q, A_i)}{N}
\end{equation}

$N$ — the total number of pairs (question, answer) in the evaluation dataset.

\subsection {Strict Binary Context Recall}

$SB CR$ evaluates whether all relevant contextual information is utilized in the answer.

% Requires: \usepackage{amsmath}
\begin{equation}
    U(C,A) = \mathds{1}\bigl(F_{\text{used}} \equiv F_{\text{relevant}}\bigr)
    \label{eq:placeholder_label3}
\end{equation}

$C,A$ — the context and the answer, respectively,

$U(C,A)$ — the binary output verdict,

$F_\text{relevant}$ — the set of all relevant facts present in the context,

$F_\text{used}$ — the set of facts from the context that are used in the answer,

% Requires: \usepackage{amsmath}
\begin{equation}
    SB\,CR = \frac{\displaystyle\sum_{i=1}^{N} U(C_i, A_i)}{N}
    \label{eq:placeholder_label4}
\end{equation}

$N$ — the total number of pairs (question, answer) in the evaluation dataset.

\subsection {Strict Binary Factual Accuracy}

$SB FA$ evaluates factual correctness and hallucination absence.
% Requires: \usepackage{amsmath}
\begin{equation}
    F(C, A) = \mathds{1}(\textit{S}(A, C) \land \lnot\textit{C}(A, C) \land \lnot\textit{H}(A, C))
    \label{eq:placeholder_label5}
\end{equation}

$C,A$ — the context and the answer, respectively,

$F(C,A)$  — the binary output verdict,

$Supported(C,A)$ — all statements in the answer are explicitly present in or directly entailed by the context,

$\lnot\textit{Contradicts}(A, C)$ — the answer does not contradict the information in the context,

$\lnot\textit{Hallucinates}(A, C)$ — the answer does not contain information that is absent from the context,

% Requires: \usepackage{amsmath}
\begin{equation}
    SB\,FA = \frac{\sum_{i=1}^{N} F(C_i, A_i)}{N}
    \label{eq:placeholder_label6}
\end{equation}

$N$ — the total number of pairs (question, answer) in the evaluation dataset.

\subsection{Retrieval Coverage Score}
$RCS$ measures how completely the retrieved context reproduces the original (reference) context and therefore characterizes the quality of the retriever.

Let $C_{orig}$ be the original context and $C_{retr}$ the context retrieved by the retriever.
The evaluation function $S(C_{orig}, C_{retr})$ is defined as follows:

\begin{equation}
S(C_{orig}, C_{retr}) = 
\begin{cases} 
    \text{0}, &
    \text{if } Cov(C_{orig}, C_{retr}) = none \\
    \text{1}, & 
    \text{if } Cov(C_{orig}, C_{retr}) = partial \\
    \text{2}, & 
    \text{if } Cov(C_{orig}, C_{retr}) = complete \\
\end{cases}
\end{equation}

Here, $Coverage(S(C_{orig}, C_{retr}))$ is either an expert assessment or the result of an LLM-based evaluator that compares the two texts.

$S = 0$ — the retrieved context $C_{retr}$ does not contain information from the original context $C_{orig}$,

$S = 1$ — the retrieved context $C_{retr}$ contains part of the information from $C_{orig}$, but not all of it,

$S = 2$ — all information from the original context $C_{retr}$ is present in the retrieved context $C_{orig}$.

The final $RCS$ value for a dataset is computed as the arithmetic mean:

\begin{equation}
    RCS = \frac{\sum_{i=1}^{N} S(C_{orig}^i, C_{retr}^i)}{N}
    \label{eq:placeholder_label7}
\end{equation}

Where $N$ is the total number of queries in the evaluation dataset.

\subsection{Domain Match Accuracy and Collection Match Accuracy}

$D\_match$ and $C\_match$ measure the structural correctness of retrieval by evaluating whether the system correctly identifies the domain and collection associated with a given query.

For each query $i$, the system produces predicted labels $d_{\text{pred}}(i)$ and $c_{\text{pred}}(i)$, while the evaluation dataset contains the corresponding reference labels $d_{\text{true}}(i)$ and $c_{\text{true}}(i)$. The metrics are defined as the proportion of correctly identified labels across the evaluation dataset.

\begin{equation}
    D_{\text{match}} = \frac{1}{N} \sum_{i=1}^{N}
    \mathbb{I}\left[d_{\text{pred}}(i) = d_{\text{true}}(i)\right]
    \label{eq:placeholder_label8}
\end{equation}

\begin{equation}
    C_{\text{match}} = \frac{1}{N} \sum_{i=1}^{N}
    \mathbb{I}\left[c_{\text{pred}}(i) = c_{\text{true}}(i)\right]
    \label{eq:placeholder_label9}
\end{equation}

where $N$ is the total number of queries, and $\mathbb{I}$ is the indicator function.

\section{Experiment}

The goal of the experiment was to evaluate the quality of information retrieval and question answering using the DCD approach on a synthetic dataset in comparison with alternative retrieval methods.

The experiment employed a simplified variation of DCD, in which an LLM classifier with structured outputs is applied before vector retrieval and sequentially identifies the target domain and collection associated with a query.
The underlying corpus was constructed from real information about residential complexes provided by a real-estate development client. Real names and numerical values were replaced with synthetic ones while preserving the original structure and relationships within the data.

The evaluated baselines included Naive RAG, Naive RAG with reranking, Contextual RAG, and RAPTOR. In addition to the baseline DCD configuration, two extensions were evaluated: DCD with cross-encoder reranking and a hybrid configuration operating on top of a contextual index.

The evaluated approaches are briefly described below.
Naive RAG is a standard retrieval-augmented generation pipeline. User queries are embedded using a single embedding model, top-k chunks are retrieved from a unified vector index, and the retrieved chunks are passed to the language model as context for answer generation.
Naive RAG + Reranker extends the baseline pipeline with a cross-encoder reranker. The top-20 retrieved candidates are reranked through full query-document interaction, and the highest-ranked top-10 chunks are used for answer generation.

Contextual RAG [\hyperlink{Anthropic}{Anthropic, 2024}] modifies the indexing stage by enriching each chunk with an LLM-generated prefix describing its position within the hierarchy (domain, collection, document). Retrieval is then performed over the enriched index and combined with reranking.
RAPTOR [\hyperlink{Sarthi}{Sarthi et al., 2024}] employs a hierarchical retrieval index. Chunks are recursively clustered, an LLM generates summaries for each cluster, and a multi-level abstraction tree is constructed. Retrieval is performed simultaneously across all hierarchy levels.

DCD is the approach investigated in this work, evaluated in a simplified configuration. A two-stage LLM classifier with structured outputs sequentially identifies the target domain and collection associated with a query. Vector retrieval is then restricted to the corresponding segment of the index. Literal-constrained outputs guarantee that the classifier always produces valid labels. Compared to the original DCD formulation, the evaluated configuration reduces the hierarchy from three levels to two and does not perform explicit document-level selection. Three configurations are considered: baseline DCD, DCD + Reranker, and DCD + Contextual.

The evaluation consisted of five stages:

\begin{itemize}
    \item Synthetic corpus generation;
    \item Generation of question--answer--context evaluation data;
    \item Construction of vector databases and auxiliary indices;
    \item Inference across all evaluated pipelines;
    \item Metric computation.
\end{itemize}

At the first stage, a synthetic corpus was generated using gpt-oss-120b based on predefined templates. The corpus consisted of fifteen domains representing residential complexes. Within each domain, five thematic collections were defined: apartments, infrastructure, management\_company, commercial\_premises, and security\_and\_access, reflecting a typical information structure in such systems. Variability was introduced through parameterized substitution of numerical values while preserving the domain–collection–document hierarchy.

At the second stage, gpt-oss-120b was used to generate question–answer–context triples for every document. Each question was prefixed with a pattern of the form "RC , section : ...", explicitly indicating the target domain and collection. This setup serves as an analogue of session-level context commonly available in real conversational scenarios. The final evaluation set contained 600 triples and was used as a common benchmark for all evaluated pipelines.

At the third stage, vector indices were constructed in ChromaDB. The base index was created by splitting documents into chunks (window size 450 tokens, overlap 30 tokens) and generating embeddings using bge-m3. Each chunk retained metadata describing its associated domain, collection, and document. These metadata were used both for evaluating retrieval correctness and for filtering within DCD. The DCD + Contextual configuration used a context-enriched index described below.

Two additional specialized indices were also constructed.
For the contextual index, qwen3.6-35b-a3b generated a one- or two-sentence contextual prefix for each chunk. The enriched text was embedded using bge-m3 and stored in a separate ChromaDB collection.

The RAPTOR index was built using RaptorPack in collapsed mode. Documents were recursively clustered, model generated summaries for each cluster, and a multi-level tree structure was constructed. All nodes were embedded using bge-m3 [\hyperlink{Chen}{Chen et al., 2024}]. The contextual index was used by both Contextual RAG and DCD + Contextual. RAPTOR relied on a separate RAPTOR index.

At the fourth stage, all seven pipelines were evaluated on the benchmark consisting of 600 queries. Query embeddings were generated using bge-m3.
For pipelines employing reranking: Naive RAG + Reranker, Contextual RAG, DCD + Reranker, and DCD + Contextual. The top-20 retrieved candidates were reranked using bge-reranker, after which the top-10 candidates were passed to the language model. In all other pipelines, the top-10 retrieved candidates were used directly.

In all DCD configurations, retrieval was preceded by a two-stage LLM router based on qwen3.6-35b-a3b with structured outputs implemented through a Pydantic schema with Literal-constrained labels. The router sequentially classified each query by domain and collection before retrieval.

Final answer generation was performed by qwen3.6 with temperature set to zero across all pipelines. Using a fixed generation model allows performance differences to be attributed primarily to retrieval quality rather than generation variability.
Finally, we computed the metrics described in Section 5.

\begin{table}[H]
\centering
\caption{Results on the synthetic evaluation dataset}
\label{tab:results}
\begin{tabular}{lcccccc}
\toprule
\textbf{Pipeline} & \textbf{SB\_ARC} & \textbf{SB\_CR} & \textbf{SB\_FA} & \textbf{RCS} & \textbf{D\_match} & \textbf{C\_match} \\
\midrule
Naive RAG & 0.655 & 0.893 & 0.733 & 1.288 & 0.56 & 0.90 \\
Naive + Reranker & 0.667 & 0.855 & 0.678 & 1.230 & 0.55 & 0.97 \\
Contextual RAG & 0.917 & 0.973 & 0.823 & 1.982 & 1.00 & 1.00 \\
RAPTOR & 0.850 & 0.940 & 0.808 & 1.827 & 0.88 & 0.92 \\
DCD & 0.885 & 0.948 & 0.965 & 1.925 & 0.96 & 0.99 \\
DCD + Reranker & 0.893 & 0.965 & 0.962 & 1.945 & 0.97 & 0.99 \\
DCD + Contextual & 0.900 & 0.947 & 0.960 & 1.912 & 0.95 & 0.99 \\
\bottomrule
\end{tabular}
\end{table}

Contextual RAG achieves the highest values on the soft evaluation metrics, reaching $SB\,ARC = 0.917$, $SB\,CR = 0.973$, and $RCS = 1.982$. However, DCD outperforms Contextual RAG on the key metric of factual accuracy. DCD achieves $SB\,FA = 0.965$, compared to $SB\,FA = 0.823$ for Contextual RAG, corresponding to an improvement of 14.2 percentage points. This result suggests improved factual consistency with the retrieved context, a desirable property for production RAG systems.

\begin{figure} [H]
    \centering
    \includegraphics[width=1\linewidth]{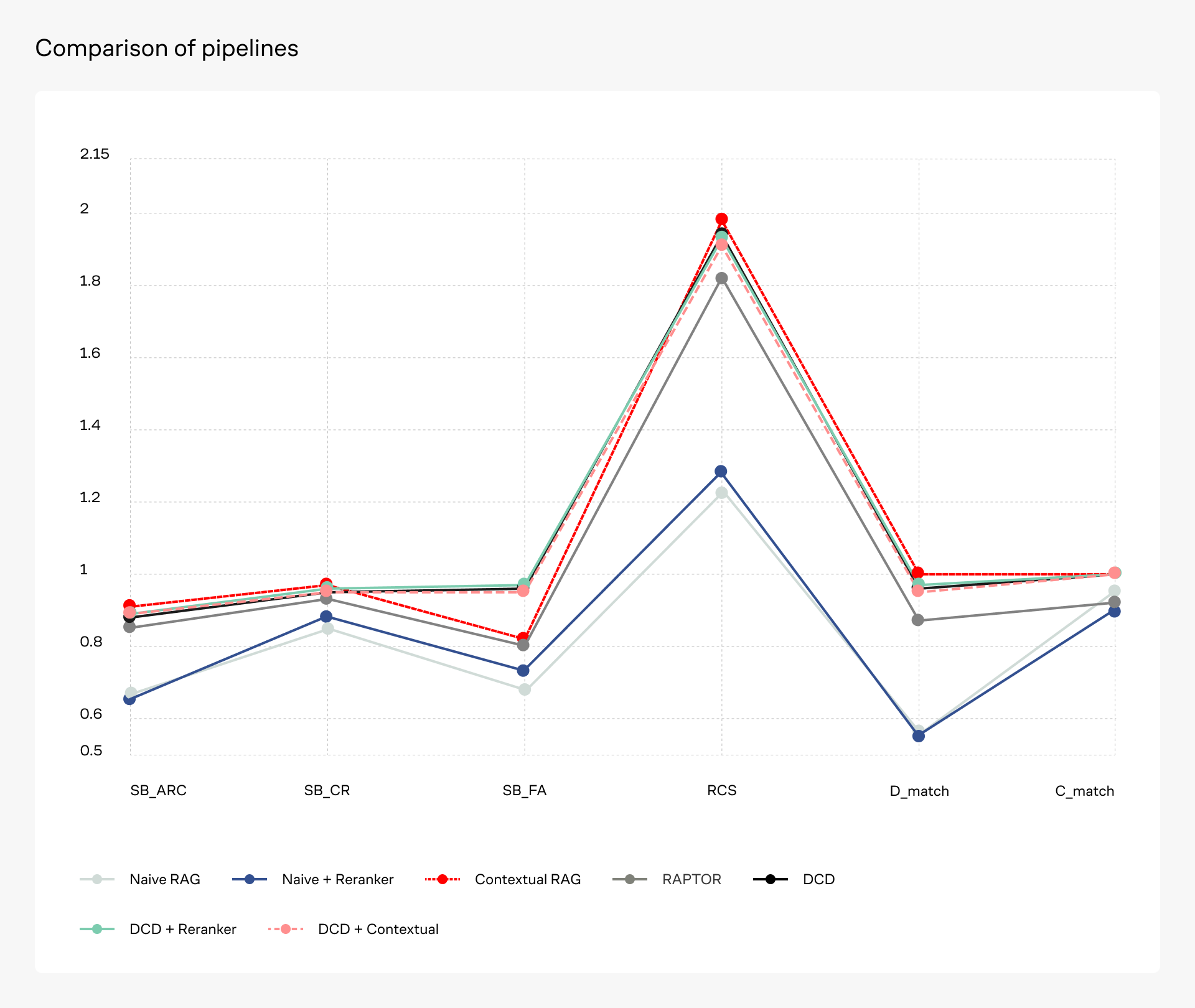}
    \caption{Comparison of evaluated pipelines}
    \label{fig:placeholder}
\end{figure}

The cost structure of the two approaches is fundamentally different. Contextual RAG requires $O(N)$ LLM calls during indexing, where $N$ denotes the number of chunks in the corpus, in order to generate contextual prefixes. Any modification of the hierarchy schema, such as adding a new field or renaming a collection, requires complete reindexing of the corpus. In contrast, DCD does not require any LLM calls during indexing and instead reuses the existing metadata hierarchy of the knowledge base. The approach incurs only two lightweight classifier calls at inference time. For corpora containing $10^{6}$--$10^{7}$ chunks and undergoing regular reindexing, this difference may become a significant operational consideration.

DCD also changes the nature of retrieval errors. Structured outputs with Literal-constrained labels guarantee that the router always returns a valid domain--collection pair, and retrieval is restricted to the corresponding segment of the index. As a result, retrieval is confined to a single domain--collection segment, preventing chunks from unrelated segments of the knowledge base from entering the context. The trade-off is the possibility of complete retrieval failure when routing is incorrect: if either the domain or collection is misclassified, the relevant chunk becomes unreachable. On the evaluation dataset, routing errors were rare. DCD achieved $D_{\text{match}} = 0.965$ and $C_{\text{match}} = 0.990$, indicating highly reliable routing performance. By contrast, in Contextual RAG and RAPTOR, incorrectly retrieved chunks may coexist with relevant ones in the final context and can only be detected indirectly through answer-quality metrics.

RAPTOR exhibits limitations on domain-segmented corpora. During recursive clustering, chunks originating from different residential complexes can be merged into shared summary nodes, causing domain-level information to be partially lost at higher levels of abstraction. This behavior is reflected in $D_{\text{match}} = 0.883$, compared to $D_{\text{match}} = 0.96$--$1.00$ for the remaining approaches, and contributes to a lower factual accuracy score ($SB\,FA = 0.808$, compared to $SB\,FA > 0.96$ for DCD). When explicit semantic boundaries are available, methods such as DCD and Contextual RAG that directly exploit these boundaries appear more effective in this setting than unsupervised clustering approaches.

\section{Limitations}

\subsection{Configuration Complexity}
The DCD approach introduces additional configuration complexity as the size and heterogeneity of the knowledge base increase. Maintaining clear semantic boundaries between domains and collections becomes more difficult as the number of knowledge areas grows, especially when adjacent domains share overlapping terminology or document structures [\hyperlink{Yao}{Yao et al., 2023}].

The experimental evaluation was conducted using a reduced routing configuration in which retrieval-space restriction is performed at the domain and collection levels through a structured-output classifier. This setup preserves the core principle of hierarchical retrieval-space restriction while reducing inference overhead and simplifying the experimental setup.

Accordingly, the limitations discussed in this section apply to the broader DCD approach and should be interpreted in the context of the evaluated configuration.

\subsection{Fully Unstructured Data}
DCD assumes the presence of at least minimal semantic organization within the knowledge corpus. When applied to fully unstructured data, the effectiveness of structured retrieval and routing mechanisms degrades, often converging toward the behavior of advanced naive RAG pipelines [\hyperlink{Gao}{Gao et al., 2021}].

\subsection{Computational Overhead}

It is important to consider the specifics of the generated dataset: all documents have a template-based structure in which texts within each domain are almost identical and differ only in specific values such as names and numerical characteristics, while preserving the same overall syntactic and semantic structure. [\hyperlink{Shazeer}{Shazeer et al., 2017}].

For this type of data, characterized by a high degree of contextual overlap, DCD predictably demonstrates higher retrieval quality compared to the naive RAG approach. 

\subsection{Evaluation Methodology}

Answer quality in this study was assessed using an LLM-as-a-Judge setup with the gpt-oss-120b model. The judge evaluated generated answers according to the evaluation metrics defined in Section 5, receiving as input the user query, the retrieved context, and the generated answer. The evaluation prompts instructed the model to assess responses according to the metric definitions and to produce a structured justification before assigning the final decision. The temperature for these evaluations was set to 0 to ensure deterministic scoring.

Retrieval quality was evaluated separately using $RCS$. In this setting, the judge compared the retrieved context with the reference context and assigned a score on a three-point scale from 0 to 2, reflecting whether the retrieved information was absent, partially sufficient, or fully matched the reference context. For $RCS$ scoring, the temperature was set to 0.1 to allow limited variability in explanatory reasoning.

In addition, two structural metrics, $D_{\text{match}}$ and $C_{\text{match}}$, were computed for each pipeline. These metrics measure the proportion of queries for which the system correctly identified the reference domain and collection.

For pipelines with explicit routing mechanisms, namely DCD and DCD + Reranker, the predictions are taken directly from the output of the LLM classifier,  whereas for all other pipelines they are inferred from the metadata of the top-ranked retrieved chunk. Unlike the answer-quality metrics, $D_{\text{match}}$ and $C_{\text{match}}$ are computed deterministically and do not rely on LLM-based evaluation.

\section{Conclusion}

We introduced DCD, a domain-oriented RAG design based on explicit knowledge hierarchies and controlled multi-stage retrieval workflows. Unlike conventional RAG pipelines that operate over a unified retrieval space, DCD constrains retrieval through hierarchical decomposition of the knowledge base and sequential routing across semantic levels. The approach is designed to leverage knowledge structures that are already present in many enterprise information systems.

To evaluate the proposed approach, we constructed a synthetic benchmark derived from real client data and compared DCD against Naive RAG, Contextual RAG, RAPTOR, and reranking-based baselines. The results demonstrate that hierarchical retrieval-space restriction can improve retrieval quality and factual consistency while maintaining predictable computational costs. In particular, DCD achieved the highest factual accuracy among the evaluated approaches and demonstrated highly reliable routing performance according to the proposed structural metrics.

The findings suggest that explicit knowledge organization can serve as an effective alternative to increasingly complex agentic and reasoning-based retrieval architectures in domains where semantic boundaries are already known and can be formalized. Rather than introducing additional reasoning loops, DCD improves retrieval robustness through controllable knowledge partitioning and constrained search.

Several limitations remain. The current study was conducted on a synthetic benchmark derived from a single application domain, and the evaluated DCD configuration employed only a two-level hierarchy without explicit document-level routing. Future work includes large-scale evaluation on heterogeneous enterprise corpora, replacing general-purpose LLM routers with specialized lightweight classifiers, and extending the approach with backward navigation and contextual relevance validation mechanisms.

\section{Resources}

Dataset: \href{https://huggingface.co/datasets/redmadrobot-rnd/dcd}{Hugging Face}

Code repository: \href{https://github.com/redmadrobot-rnd/dcd}{GitHub}

Both resources are maintained by the AI R\&D team at red\_mad\_robot.

If you have any questions about this work, please contact us at \texttt{rnd@redmadrobot.com}.

\section{References}

%Здесь мы создаем \hypertarget{my_anchor}{«Секретную точку»}. Запомните это место!
%\hyperlink{my_anchor}{прыгнуть к секретной точке}.

\noindent\hangindent=0.5cm\hangafter=1
\hypertarget{Anthropic}{Anthropic.}
\textit{Introducing Contextual Retrieval}. Technical report / engineering blog, 2024.

\noindent\hangindent=0.5cm\hangafter=1
\hypertarget{Chen}{Chen, J.}, Xiao, S., Zhang, P., Luo, K., Lian, D., Liu, Z.
\textit{M3-Embedding: Multi-Linguality, Multi-Functionality, Multi-Granularity Text Embeddings Through Self-Knowledge Distillation}. Findings of ACL, 2024.

\noindent\hangindent=0.5cm\hangafter=1
\hypertarget{Gao}{Gao, L.}, Ma, X., Lin, J., Callan, J.
\textit{Precise Zero-Shot Dense Retrieval without Relevance Labels}. ACL, 2021.

\noindent\hangindent=0.5cm\hangafter=1
\hypertarget{Guu}{Guu, K.}, Lee, K., Tung, Z., Pasupat, P., Chang, M.-W.
\textit{REALM: Retrieval-Augmented Language Model Pre-Training}. International Conference on Machine Learning (ICML), 2020.

\noindent\hangindent=0.5cm\hangafter=1
\hypertarget{Izacard}{Izacard, G.}, Grave, E.
\textit{Leveraging Passage Retrieval with Generative Models for Open Domain Question Answering}. European Chapter of the ACL (EACL), 2021.

\noindent\hangindent=0.5cm\hangafter=1
\hypertarget{Ji}{Ji, Z.}, Lee, N., Frieske, R., et al.
\textit{Survey of Hallucination in Natural Language Generation}. ACM Computing Surveys, 2023.

\noindent\hangindent=0.5cm\hangafter=1
\hypertarget{Khot}{Khot, T.}, Sabharwal, A., Clark, P.
\textit{Decomposed Prompting: A Modular Approach for Solving Complex Tasks}. International Conference on Learning Representations (ICLR), 2023.

\noindent\hangindent=0.5cm\hangafter=1
\hypertarget{Lewis}{Lewis, P.}, Perez, E., Piktus, A., et al.
\textit{Retrieval-Augmented Generation for Knowledge-Intensive NLP Tasks}. Advances in Neural Information Processing Systems (NeurIPS), 2020.

\noindent\hangindent=0.5cm\hangafter=1
\hypertarget{Li}{Li, Y.}, Zhang, W., Yang, Y., et al.
\textit{Towards Agentic RAG with Deep Reasoning: A Survey of RAG-Reasoning Systems in LLMs}. arXiv preprint, 2025.

\noindent\hangindent=0.5cm\hangafter=1
\hypertarget{LiuGEval}{Liu, Y.}, Iter, D., et al.
\textit{G-Eval: NLG Evaluation using GPT-4 with Better Human Alignment}. arXiv preprint arXiv:2303.16634, 2023.

\noindent\hangindent=0.5cm\hangafter=1
\hypertarget{Lin}{Lin, S.}, Hilton, J., Evans, O., et al.
\textit{TruthfulQA: Measuring How Models Mimic Human Falsehoods}. ACL, 2022.

\noindent\hangindent=0.5cm\hangafter=1
\hypertarget{Makin}{Makin, A.}
\textit{Ontology-Driven Knowledge Management Systems Enhanced by Large Language Models}. ResearchGate preprint, 2024.

\noindent\hangindent=0.5cm\hangafter=1
\hypertarget{Manakul}{Manakul, P.}, Liusie, A., Gales, M.
\textit{SelfCheckGPT: Zero-Resource Black-Box Hallucination Detection for Generative Large Language Models}. EMNLP, 2023.

\noindent\hangindent=0.5cm\hangafter=1
\hypertarget{Nair}{Nair, I.}, Somasundaram, S., Saxena, A., Goswami, K.
\textit{Drilling Down into the Discourse Structure with LLMs for Long Document Question Answering}. arXiv preprint, 2023.

\noindent\hangindent=0.5cm\hangafter=1
\hypertarget{OpenAI}{OpenAI.}
\textit{Structured Outputs in Language Models}. Technical report / documentation, 2024.

\noindent\hangindent=0.5cm\hangafter=1
\hypertarget{Sarthi}{Sarthi, P.}, et al.
\textit{RAPTOR: Recursive Abstractive Processing for Tree-Organized Retrieval}. arXiv preprint, 2024.

\noindent\hangindent=0.5cm\hangafter=1
\hypertarget{Shazeer}{Shazeer, N.}, Mirhoseini, A., Maziarz, K., et al.
\textit{Outrageously Large Neural Networks: The Sparsely-Gated Mixture-of-Experts Layer}. ICLR, 2017.

\noindent\hangindent=0.5cm\hangafter=1
\hypertarget{Wei}{Wei, J.}, Wang, X., Schuurmans, D., et al.
\textit{Chain-of-Thought Prompting Elicits Reasoning in Large Language Models}. NeurIPS, 2022.

\noindent\hangindent=0.5cm\hangafter=1
\hypertarget{Yao}{Yao, S.}, Zhao, J., Yu, D., et al.
\textit{Tree of Thoughts: Deliberate Problem Solving with Large Language Models}. arXiv preprint arXiv:2305.10601, 2023.

\noindent\hangindent=0.5cm\hangafter=1
\hypertarget{Zheng}{Zheng, L.}, Chiang, W.-L., Sheng, Y., et al.
\textit{Judging LLM-as-a-Judge with MT-Bench and Chatbot Arena}. NeurIPS, 2023.

\end{document}